\documentclass[10pt,journal,compsoc]{IEEEtran}

\ifCLASSOPTIONcompsoc
  \usepackage[nocompress]{cite}
\else
  \usepackage{cite}
\fi

\ifCLASSINFOpdf
\else
\fi
\usepackage{amssymb}
\usepackage{bm}
\usepackage{enumerate}
\usepackage{url}
\usepackage{rotating}
\usepackage{listings}
\usepackage{multicol}
\usepackage{changes}

\usepackage{pgffor}
\usepackage{mathtools, cuted}

\usepackage{multirow}
\usepackage{tikz,pgfplots}
\usepackage{subfigure}
\usepackage{units}
\usepackage{ifthen}
\usepackage{enumerate}
\usepackage[shortlabels]{enumitem}

\usetikzlibrary{calc}

\usepackage{amsmath}

\newcommand{\balpha}{\mbox{\boldmath$\alpha$}}

\usepackage{array}
\usepackage{hyperref} 
\newcolumntype{L}{>{\centering\arraybackslash}m{1.5cm}}

\usepackage{scalerel}
\usepackage{tikz}
\usetikzlibrary{svg.path}

\definecolor{orcidlogocol}{HTML}{A6CE39}
\tikzset{
  orcidlogo/.pic={
    \fill[orcidlogocol] svg{M256,128c0,70.7-57.3,128-128,128C57.3,256,0,198.7,0,128C0,57.3,57.3,0,128,0C198.7,0,256,57.3,256,128z};
    \fill[white] svg{M86.3,186.2H70.9V79.1h15.4v48.4V186.2z}
                 svg{M108.9,79.1h41.6c39.6,0,57,28.3,57,53.6c0,27.5-21.5,53.6-56.8,53.6h-41.8V79.1z M124.3,172.4h24.5c34.9,0,42.9-26.5,42.9-39.7c0-21.5-13.7-39.7-43.7-39.7h-23.7V172.4z}
                 svg{M88.7,56.8c0,5.5-4.5,10.1-10.1,10.1c-5.6,0-10.1-4.6-10.1-10.1c0-5.6,4.5-10.1,10.1-10.1C84.2,46.7,88.7,51.3,88.7,56.8z};
  }
}

\newcommand\orcidicon[1]{\href{https://orcid.org/#1}{\mbox{\scalerel*{
\begin{tikzpicture}[yscale=-1,transform shape]
\pic{orcidlogo};
\end{tikzpicture}
}{|}}}}

\begin{document}
This work has been submitted to the IEEE for possible publication. Copyright may be transferred without notice, after which this version may no longer be accessible.

%
\title{Fog Device-as-a-Service (FDaaS): A Framework for Service Deployment in Public Fog Environments}
%
%
%
%

\author{Sudheer Kumar~Battula ~\orcidicon{0000-0001-6597-252X},~\IEEEmembership{Member,~IEEE,}
        Saurabh~Garg~\orcidicon{0000-0003-3510-2464},~\IEEEmembership{Member,~IEEE,}
        James~Montgomery~\orcidicon{0000-0002-5360-7514},~\IEEEmembership{Member,~IEEE,}
        and~Ranesh~Naha~\orcidicon{0000-0003-4165-9349},~\IEEEmembership{Member,~IEEE}
        \IEEEcompsocitemizethanks{\IEEEcompsocthanksitem S. K. Battula, S. Garg, J. Montgomery and R. K. Naha is with the School of Information and Communication Technology, University of Tasmania, TAS 7005, Australia,
\protect\\
E-mail: \{sudheerkumar.battula; saurabh.garg; james.montgomery; raneshkumar.naha\}@utas.edu.au \\
Corresponding author: S. K. Battula}
\thanks{Manuscript received March XX, 2021; revised XXXXX XX, 2021.}}

\markboth{Journal of \LaTeX\ Class Files,~Vol.~14, No.~8, August~2020}%
{Shell \MakeLowercase{\textit{et al.}}: Bare Advanced Demo of IEEEtran.cls for IEEE Computer Society Journals}
%



\IEEEtitleabstractindextext{%
\begin{abstract}
Meeting the requirements of future services with time sensitivity and handling  sudden load spikes of the services in Fog computing environments are challenging tasks due to the lack of publicly available Fog nodes and their characteristics. Researchers have assumed that the traditional autoscaling techniques, with lightweight virtualisation technology (containers), can be used to provide autoscaling features in Fog computing environments, few researchers have built the platform by exploiting the default autoscaling techniques of the containerisation orchestration tools or systems. However, the adoption of these techniques alone, in a publicly available Fog infrastructure, does not guarantee Quality of Service (QoS) due to the heterogeneity of Fog devices and their characteristics, such as frequent resource changes and high mobility. To tackle this challenge, in this work we developed a Fog as a Service (FaaS) framework that can create, configure and manage the containers which are running on the Fog devices to deploy services. This work presents the key techniques and algorithm which are responsible for handling sudden load spikes of the services to meet the QoS of the application. This work was provided an evaluation by comparing it with existing techniques under real scenarios. The experiment results show that our proposed approach maximises the satisfied service requests by an average of 1.9 times in different scenarios.

\end{abstract}

 \begin{IEEEkeywords}
 Fog Computing, Resource Deployment, Autoscaling, and Internet of Things.
 \end{IEEEkeywords}}

 \maketitle



\section{Introduction}
\label{sec:7introduction}

\IEEEPARstart{F}{og} computing is the latest computing paradigm that brings computation power closer to the end devices to provide services similar to the Cloud for a wide range of smarter applications \cite{ujjwal2019cloud,ujjwal2020cloud,tammishetty2017iot,tuli2020healthfog}. In public settings, anyone can participate in the Fog computing environment by providing devices for processing the service requests of the users. This infrastructure is not fully dedicated to providing services and there is no centralised control. Fog Systems are required to scale up the resources by deploying the services in the situation in which the number of users requesting the services is often growing and will continue to rise which leads to an increase in the  time to serve requests.

The process of deploying and managing services in this volunteer environment, in order to efficiently handle the sudden load spikes of the services, needs to find the exact number of suitable devices to scale and descale the virtual Fog devices. In this way Quality of Service (QoS) is ensured while maximising overall resource utilisation. If the system estimates the number of devices to scale to be low, it will lead to under-provisioning. The load on the devices will increase and could fail to meet the QoS of the applications. If the system estimates the number of devices to scale to be high, it will lead to over-provisioning. Due to the scarcity of resources, the number of rejections of the service requests may be high and the cost increased. Hence, this process is a complex task due to the uncertain behaviour, high dynamics of resource changes and high mobility of Fog devices in Fog computing environments. Moreover, due to the fact that services are subscribed by a large number of end-users, the demand of the application load can spike at any point in time.

There are a few Fog frameworks, platforms and services that exist to manage Fog devices and orchestrate the application workload. Santoro et al.~\cite{santoro2017foggy} proposed a workload orchestration architectural framework and platform called Foggy. Powell et al.  \cite{8957674} proposed a Fog Development Kit (FDK) that abstracts the complexities of resource allocation and provides high-level interfaces to the developers for rapid development of the systems. However, the authors considered that autoscaling feature as their future work.

A number of researchers have built Fog systems using existing container orchestration and analytical tools; they provide autoscaling features by exploiting the default autoscaling scheme, technique or feature of the orchestration tools~\cite{hong2016dynamic,tsai2017distributed,zheng2018auto,kayal2020kubernetes}. However, in these default schemes, the resources are scaled, based on the currently available resources. Therefore, in these schemes, the fact that the resources are not fully dedicated to providing services was ignored. Thus, the adoption of these schemes or techniques alone in a volunteer Fog infrastructure does not guarantee the Quality of Service (QoS) of services.

Thus, this work aims to develop the Fog as a service that can deliver effective and autoscaling IoT services in a Fog environment. The main idea is to develop a framework that can efficiently provide ondemand virtual Fog resources for IoT and other services even when there are sudden load spikes. 

The contributions of this paper are as follows:
\begin{enumerate}

\item To propose and implement a FaaS to deploy services efficiently in public Fog environments.
\item An autoscaling technique based on the prediction of future workloads and dynamically estimating the future state of the Fog devices based on the current state.
\end{enumerate}

The paper is organised as follows. In Section \ref{sec:7RelatedWork}, related works in the context of autoscaling in Fog computing systems and platforms are discussed. Section \ref{sec:7FaaSobj} describes the design objectives of the Fog as a Service. Section \ref{sec:7sysmodel} provides an overview of system model and also discusses the dynamic resource scaler algorithm  by exploiting device minimum availability information. Section \ref{sec:7performanceeval} provides detailed information about real-time cases. In Section \ref{sec:7results}, the results of the experiments obtained in different scenarios are presented. Finally, Section \ref{sec:7conclusion} concludes the paper.

\section{Related Work}
\label{sec:7RelatedWork}

Autoscaling problem has been widely studied in the area of Cloud computing and IoT. However, the techniques that are used to solve the problem are not applicable in Fog computing environments due to their unique characteristics. Moreover, to date, various platforms, emulators and frameworks have been developed to address the autoscaling problem in Fog computing environments. In this section, we will summarise the works that are closely related to the autoscaling problem in Fog computing environments.

Mayer et al. \cite{mayer2017emufog} proposed an extensible and scalable Fog emulator framework called EmuFog. In this work, the authors highlighted the key components of the framework and explained the Backbone connection algorithm and node placement algorithm. However, the authors have not considered the resource dynamics of Fog nodes and implementing mobility models as a plan. De Alfonso et al. \cite{de2017container} showed the feasibility of adopting containers in the creation of virtual computing clusters to meet the needs of the scientific applications. Authors used the open-source tool Elastic Cluster for Docker (EC4Docker) and to enable the autoscaling feature, authors integrated with Docker swarm. However, in this work, there is a manual process involved in the creation or generation of images and Docker files. Moreover, the clusters are deployed in bare metal machines.

There are many frameworks proposed to support autoscaling in Fog computing environments which exploit the scalability feature from existing container orchestration platforms. Bellavista et al. \cite{bellavista2017feasibility} proposed a Fog framework by using Kura gateway. It is Docker-based containerisation for IoT applications that suits the resource-limited Fog devices. A local broker is integrated with the Kura framework to enable autoscaling. A general Kura gateway sends sensors’ data to the Message Queue Telemetry Transport (MQTT) broker after aggregating all gathered data. The MQTT broker hosted on the Cloud manages the computation with the geographically distributed global Cloud. The proposed extension of the Kura framework completely supports the management of infrastructure such as update, download and administrating virtualised Docker images. Containerisation supports portability and interoperability which help to build the container-based Fog environment with different Fog nodes. Similarly there are other frameworks which exploit the autoscaling from Kubernetes \cite{goethals2020adaptive}\cite{rossi2020geo}\cite{9144934}. These frameworks did not consider that devices are not completely dedicated to processing the applications (public Fog devices) for autoscaling.

In reference to dynamic resource provisioning frameworks and techniques, we can also see some interesting contributions for autoscaling. A novel Fog-based communications scheme for mobile IoT crowdsensing applications called MIST has been proposed to support different workloads of IoT sensing applications based on consumer data, task distribution and virtual machine placement~\cite{arkian2017mist}. In this work, the author considers the autoscale feature as future work. Similarly, Tseng et al. \cite{8272512}  proposed a lightweight Fog platform by integrating the hypervisor and containerisation technology; they proposed a fuzzy-based autoscaling mechanism based on the current status of the resource in order to improve the scalability service of industrial applications. Yousefpour \cite{yousefpour2019fogplan} proposed a lightweight QoS-aware Dynamic Fog Service Provisioning (QDFSP) framework for Fog services called FOGPLAN. The authors formulated the deployment and release of the Fog services problem into an optimisation problem. To tackle this, the authors proposed two efficient greedy algorithms. In this work, authors consider the binary variable for decision making for deploying IoT services in the Fog or Cloud. Moreover, the authors considered the performance of horizontal and vertical scaling, based on the load variations with non-binary variables in their future work. Similarly, there are a few other service placement frameworks~\cite{nabavi2021tractor,9351960,8894519,9189841}. However, most of these works deal with service deployment or placement by either context, throughput or resource awareness. Moreover, these works consider latency and current resource utilisation as being the only factors in decision making. Since Fog nodes are volunteered devices and limited resources, we must consider the other parameters to satisfy QoS requirements of applications.
Thus, we developed a FaaS framework to fill the gap in the literature for an autoscaling framework for service deployment in public Fog environments.

   \begin{figure*}[h]
	\centering
	\includegraphics[width=\textwidth]{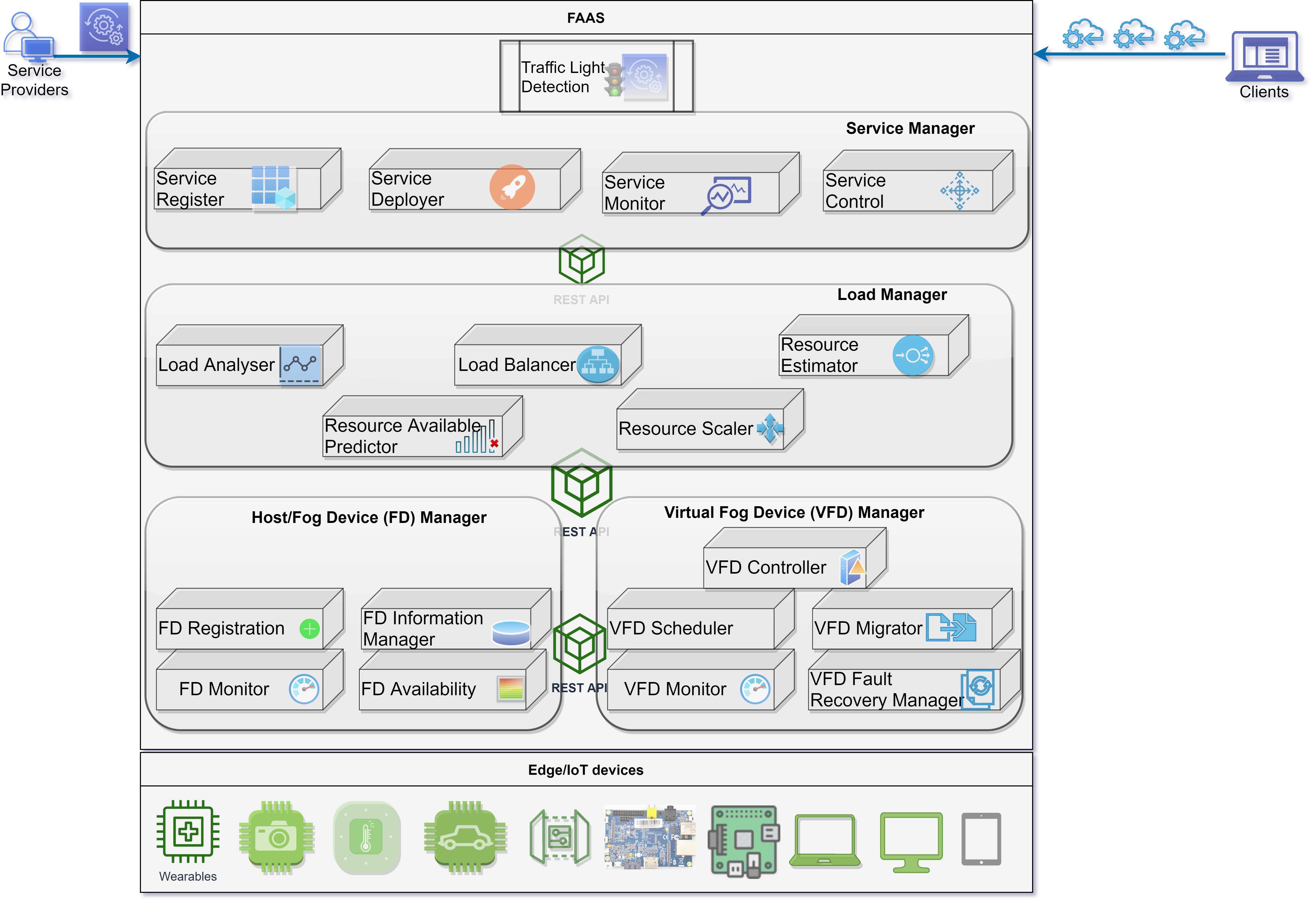}
	\caption{FAAS Framework.}
	\label{fig_Foaas}
\end{figure*}
\section{FaaS Objectives}
\label{sec:7FaaSobj}

This section describes the design objectives that are derived from the limitations and future works of the related works. This section also discusses in detail the autoscaling algorithm for scaling the nodes to meet the demands of the users/applications.

\subsection{FaaS Objectives}
 To meet the requirements of future long-running services should have the following features:

\begin{enumerate}

\item \textbf{Mobility Management}: Most of the Fog nodes in Fog computing are mobile in nature. FaaS should handle the mobility of Fog nodes automatically with the minimal number of migrations between the Fog devices.

\item \textbf {Large Scale Deployment}: Similar services must be able to run in multiple locations and each service may require multiple devices to meet the requirements of the user. FaaS makes it easier by automatically provisioning the resources based on the requirements of the applications to deploy them. 

\item \textbf{Single Click Management}: As the majority of devices are decentralised and highly mobile in nature, managing these devices physically is a complex task. FaaS is able to configure and monitor the current state of Fog devices with a single click.

\item \textbf{Resource Dynamics Management:} Fog devices in the Fog computing environments are not completely dedicated to process the requests of the user. Therefore, the resource of the devices can be changed at any time which may lead to the unavailability of the application. FaaS should handle the resource dynamics with a minimal number of failures as a result of resource scarcity.

\item \textbf{Autoscaling Management :} FaaS enables autoscaling functionality to meet the requirements of the users.

\item \textbf{Extensibility:} FaaS provides REST API that helps to modify the inbuilt techniques or reset the default parameters according to their use case scenario. 
\end{enumerate}

\section{System Models}
\label{sec:7sysmodel}

In this section, we discuss the overview of Fog as a service (FaaS) and application models. We assume that the applications or services are deployed and running within the containers to enable resource-sharing on Fog devices. We assume that in this work, the services are running continuously on containers of Fog devices that are non-dedicated and mobile in nature. Thus, to run services continuously requires multiple devices at a time.
\subsection{FaaS Model}
Fog computing provides services similar to the Cloud near the end-user. One of the services is virtualised Fog devices in the form of containers called Fog as a Service (FaaS). FaaS provides storage, computer and network in the form of containers. The FaaS model provides VFD to the users or applications to run their services in a Fog computing environment. Based on the load of the applications, the number of Fog devices can expand or shrink to satisfy the requirements of the application. Figure \ref{fig_Foaas} shows an overview of the proposed framework. Different individual manager components are responsible for managing the Fog devices, Virtual Fog devices and the application load. At the top level, the Load management component comprises different functionalities of the proposed system, such as load analyser, load balancer, resource estimator and scaler. The next level of the proposed FAAS has two components (i) Host/Fog device (FD) manager and (ii) Virtual Fog device manager. These managers can be deployed on different nodes and they communicate through the RESTful APIs. More detailed information about the managers will be explained in the following sections.


\begin{figure}[h]
	\centering
	\includegraphics[width=0.5\textwidth]{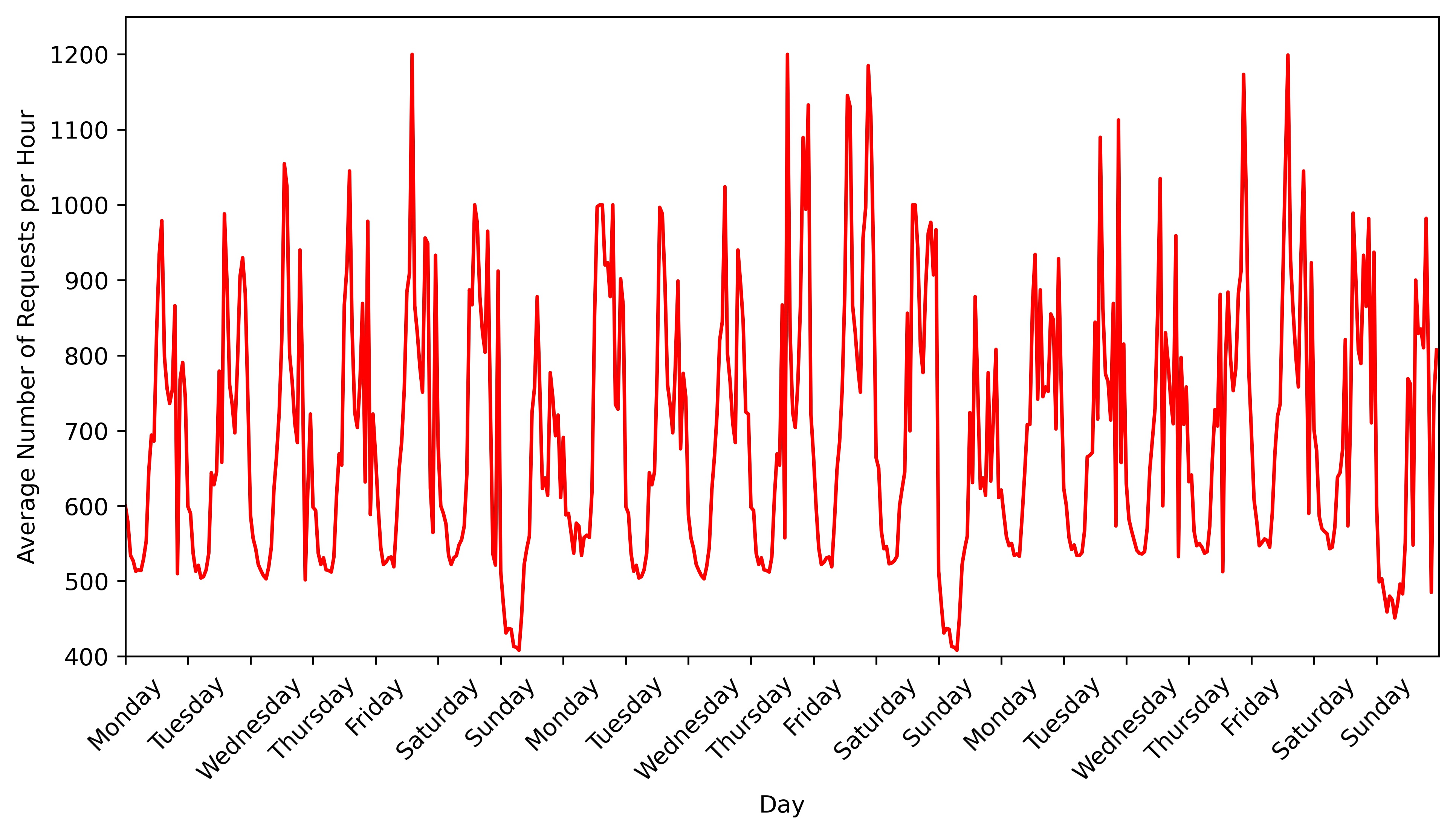}
	\caption{Historical data}
	\label{hist}
\end{figure}
\begin{figure}[h]
	\centering
	\includegraphics[width=0.5\textwidth]{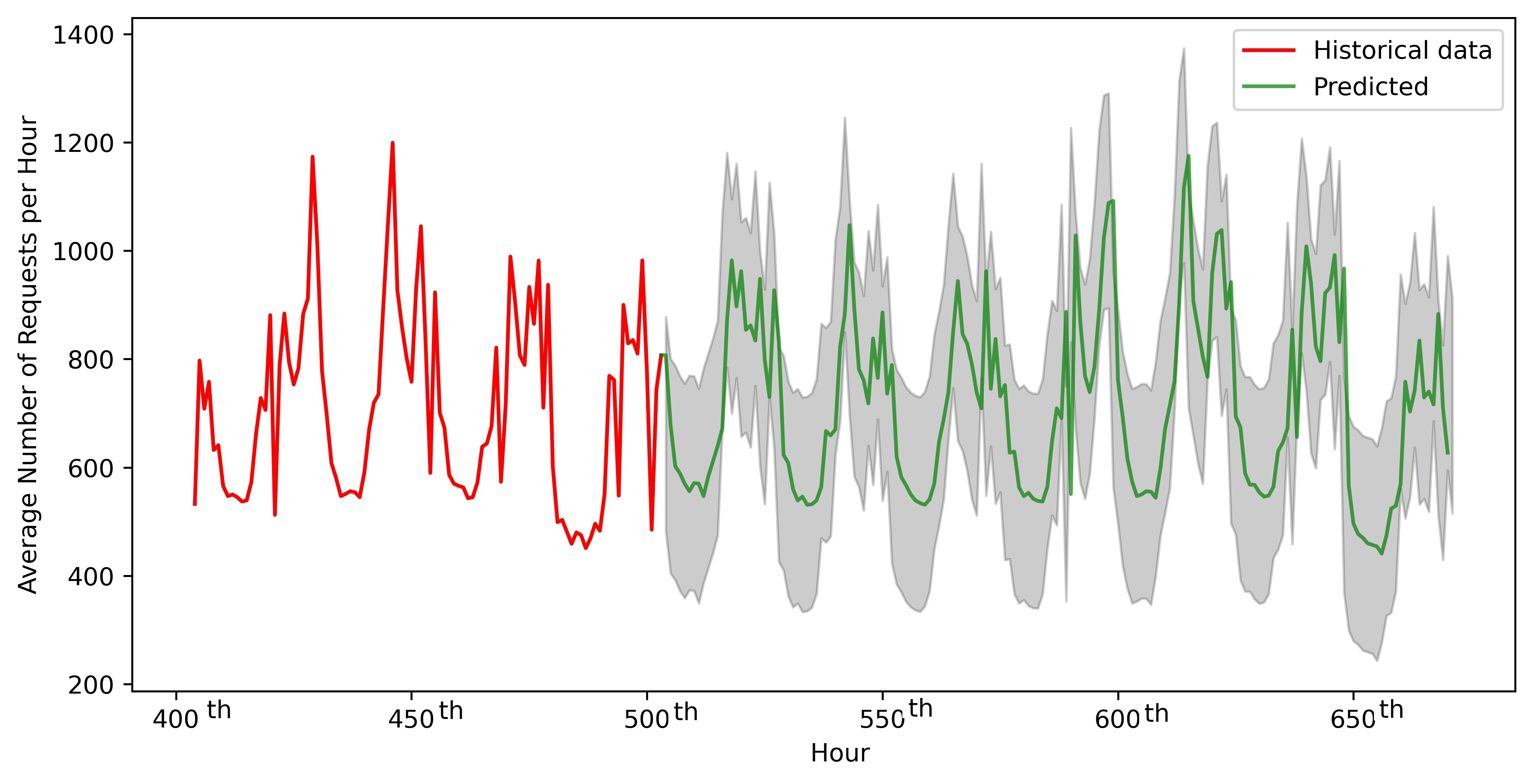}
	\caption{Predicted requests}
	\label{pred}
\end{figure}

The Load Management component has five sub-components: (i) Application Load Analyser (ii) Resource Estimator/Predictor (iii) Resource Available Predictor (iv) Dynamic Resource Scaler and (v) Load Balancer.
\textbf{Application Load analyser }:
 The Application Load Analyser yser is responsible for predicting the number of requests that can arrive at a particular time. This information is predicted based on the historical data which is collected by the application agent. Figure \ref{hist} presents the average number of requests received per hour during a three week simulations based on the workload requests rate ~\cite{calheiros2011virtual}. The future load of the application information will be sent to the resource provisioning module to estimate the number of virtual Fog devices required to scale or descale. To predict the workloads Time Series Analysis (TSA) and Time Series Modelling (TSM) are widely used. The old workload data will not have much effect on future predictions due to the uncertainty of workloads. Thus the system will store only the latest N days’ data  \cite{7943364}.We assume that the application load will follow some trend during peak times in a day. Thus, we employed the Seasonal Auto Regressive Integrative Moving Average (SARIMA) model. The model of SARIMA is equivalent to that of ARIMA, in SARIMA the seasonality of the data is taken into account. These model have two parts (i) Non-seasonal model ARIMA to predict the long term changes and (ii) Seasonal model ARIMA to predict the seasonal cycle. This is given as follows:
\begin{equation*}
    ARIMA(p,d,q) \times (P,D,Q)_{s}
\end{equation*}

where $p$, $p$ and $q$ are the non-seasonal AR order, differencing and MA order respectively. Similarly, P,D and Q are the seasonal AR order, differencing and MA order respectively. s indicates the length of seasonal data. A seasonal ARIMA model can be expressed as follows:

\begin{equation}
\phi_{p}(L)\Phi_{P}(L)^{s}\Delta^{d}\Delta_{s}^{D}y_{t}= A(t)+\Theta_{q}(L)\widetilde{\Theta_{q}}(L^{s})\varepsilon_{t}
\end{equation}

where $L$ is the backshift operator or Lag operator, $s$ is the time span of seasonal trend as we are expecting the seasonal day on an hourly basis each day s=24. $\phi_{p}$  is AR , $Phi_{P}$ is a seasonal Auto Regression (AR). $\Delta^{d}$ indicates differencing,  $\Delta_{s}^{D}$ indicates the seasonal differencing. $y_{t}$ is time series. $A(t)$ is constant trend , $\Theta_{q}(L)$ is 
moving average (MA). $\widetilde{\Theta_{q}}(L^{s})$ is seasonal MA and $\varepsilon_{t}$ is error. The workload analyser will predict the information about the average number of arrival requests on an hourly basis. The predicted workload of the SARIMA model for the above historical data is shown in Figure~\ref{pred}.   

\textbf{Resource Estimator/Predictor}: The resource predictor is responsible for estimating the number of virtual Fog devices and the type required to meet the QoS of the application. The considered service is a CPU and a memory intensive service because the images will be uploaded to detect the traffic signal. Each virtual Fog device type has a predefined size. The optimal Resource Predictor is required to estimate the number of virtual Fog devices with their type based on the future application load. We defined the Resource Predictor in the following linear form:
\begin{equation}
Q_{(t)}=\sum_{i=1}^{k} \alpha_{i}x(t_{(i)},\mu)
\end{equation}

where $Q_{(t)}$,  is the average response time at time $t$ which is specified by one of the QoS parameters. $k$ is the number of virtual Fog device types. $\alpha$  represents the number of resources of the particular type required to meet the QoS of the application.  $x(t_{(i)},\mu)$ is the lookup table that gives information about the average response time of individual type virtual Fog devices on dynamic workloads. $\mu$  is the difference between the predicted incoming application requests $IR$ and the number of requests that can still fit in the system at a time $nSRF$.
\begin{equation}
    \mu=IR-nSRF
\end{equation}

The main objective of this resource predictor is to minimise the number of virtual Fog devices which can optimally fit the requests in the virtual Fog devices to meet the QoS targets. We extract the feasible resource predictor that minimises the number of virtual Fog devices and their size, by solving the following optimisation problem:
\begin{eqnarray*}
min \sum_{i=1}^{k} \alpha_{i}x(t_{(i)},\mu)\\
\mathit{s.t.}\\
\sum_{i=1}^{k} \alpha_{i}x(t_{(i)},\mu) \le Q_{(t)}\\
0 \le \alpha_{i} \le A_{i}\\  \forall i \in 1,2,..,k\\
\mu > 0
\end{eqnarray*}

Where $A_{i}$  is the available number of resources of type i during the time interval $t \to t + \delta t$.

\textbf{Resource Available Predictor}: The resource available Predictor is responsible for predicting  $A_{i}$. The number of devices of type i is available during the time interval $t \to t + \delta t$. This can be obtained from the following equation which is from our previous work~\cite{9253552}.
	 \begin{multline}
	    	 x^*=\min \{x:  	Pr(X_{(\ell)}(u)\in {\mathcal{A}}(\ell,k,x)\\
	 \mbox{ for all }u\in[t,t+\Delta t]) \geq y\} 
	 \end{multline}

\begin{eqnarray}
\lefteqn{Pr(X_{(\ell)}(u)\in \mathcal{A}(\ell,k,x)\mbox{ for all }u\in[t,t+\Delta t])}
\label{eq_q1b}\nonumber\\
&=&\sum_{i,j\in \mathcal{A}(\ell,k,x)}
[\balpha e^{{\bf Q}_{(\ell)}t}]_i
[e^{{\bf Q}_{(\ell);\mathcal{A}(\ell,k,x)}\Delta t}]_{ij}
\label{eqq_q1b},
\end{eqnarray}
where ${\bf Q}_{(\ell);\mathcal{A}(\ell,k,x)}=[\ [{\bf Q}_{(\ell)} ]_{ij}\ ]_{i,j\in\mathcal{A}(\ell,k,x)}$ is a non-conservative (with some negative row sums) generator restricted to  transitions between the states in  $\mathcal{A}(\ell,k,x)$ only.\\

\begin{figure}[h]
	\centering
	\includegraphics[width=0.5\textwidth]{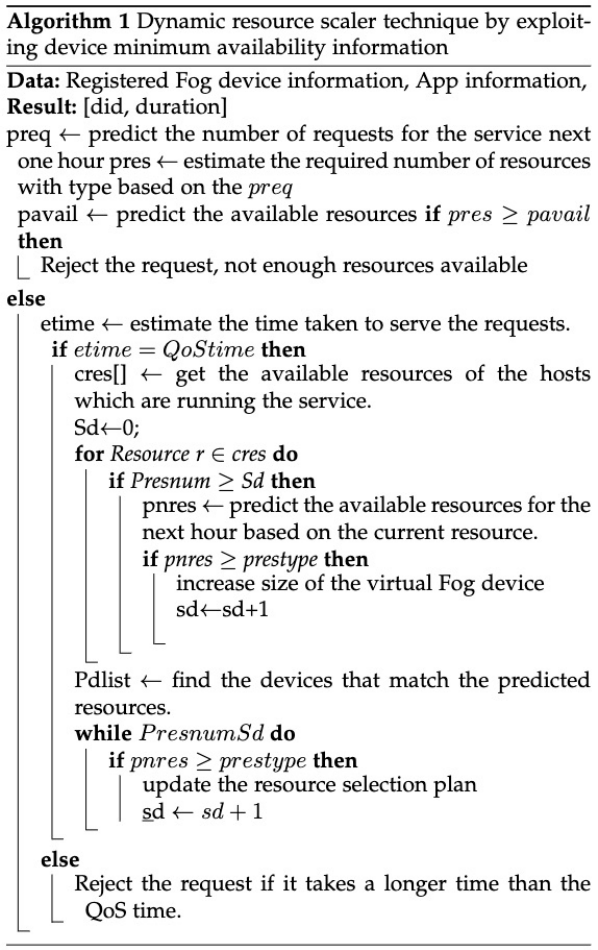}
	\label{drs}
\end{figure}

\textbf{Dynamic Resource Scaler}: 
Once the predicted resources are available to scale, the admission control mechanism will forward this request to the dynamic scaler. The dynamic resource scaler will select the required resources. Algorithm \ref{drs} provides detailed information on dynamic resource scaler service. This service checks the number of predicted resources $pres$ with the available resources $pavail$. If the  $pres$ is greater than the $pavail$ then the request will be rejected due to not enough resources available else the service will estimate the time is taken to serve the requests $etime$ and checks with the $QoStime$ if it not matches the request will be rejected due to longer time than the QoS time. Otherwise, gets the host machine list $cres$ which is already running the services. Check each host machine resource utilisations, such as memory, CPU and storage which are presently running the application service to decide whether to perform the vertical scaling or not. Even if the hosted machines have enough configuration to scale the resources, the dynamic resource scaler will not initiate the scale in operation because the machines are not completely dedicated to processing this service. So service will predict the available resources $pnres$ for the next hour, based on the current resource state and the available device history. If the predicted host state matches with the predicted resource type $prestype$, then it will increase the size of the VFD. This will continue with all the machines that are already hosting the application. If the predicted number of resources are still yet to scale, then it predicts and finds the devices that matched with the predicted resources $Pdlist$ and update the resource selection plan. The resource selection plan consists of $did$ which is the unique device id and duration that the device needs to allocate.



\textbf{Load Balancer:} To distribute the requests among the virtual Fog devices to meet the time-sensitive requirements of the application, we used a weighted round-robin load balancing algorithm. In this, based on the size and average processing requests per second of each virtual Fog device, we assign the weights to the machine. The machine with a higher weight will receive a higher proportion of requests.

The VFD manager component has five sub-components (i) VFD controller (ii) VFD Scheduler (iii) VFD Migrator (iv) VFD Monitor and (v) VFD Fault Recovery Manager. \\
\textbf{VFD Controller}: The virtual fog device controller is responsible for creating, starting, stopping and deleting a virtual Fog device in host devices.\\
\textbf{VFD Scheduler}: The virtual fog device scheduler is responsible for assigning host devices to VFD.\\
\textbf{VFD Migrator}: The virtual fog device migrator is responsible for migrating the VFDs to the other available and running hosts.\\
\textbf{VFD Monitor}:The virtual fog device monitor is responsible for monitoring the CPU, RAM and network usage of VFDs. \\
\textbf{VFD Fault Recovery Manager}: The virtual fog device fault recovery manager is responsible for running similar services of a failed virtual Fog device in the available Fog device. \\

The FD manager component has four sub-components (i) FD registration (ii) FD information manager  (iii) FD monitor and (iv) FD availability  \\
\textbf{FD registration}: The FD registration is responsible of the FD registration. The device is registered with all the device's configuration details and the possible mobility locations. The FD registration creates and assigns a unique identifier to identify the device.\\
\textbf{FD information manager}: The FD information manager is responsible for storing the host devices information.\\
\textbf{FD monitor}: The FD monitor is responsible for monitoring the FDs resource utilisation and location information.\\
\textbf{ FD availability}: The FD availability is responsible for maintaining the availability information of each Fog devices\\

  \begin{figure*}[t]
	\centering
	\includegraphics[width=\textwidth]{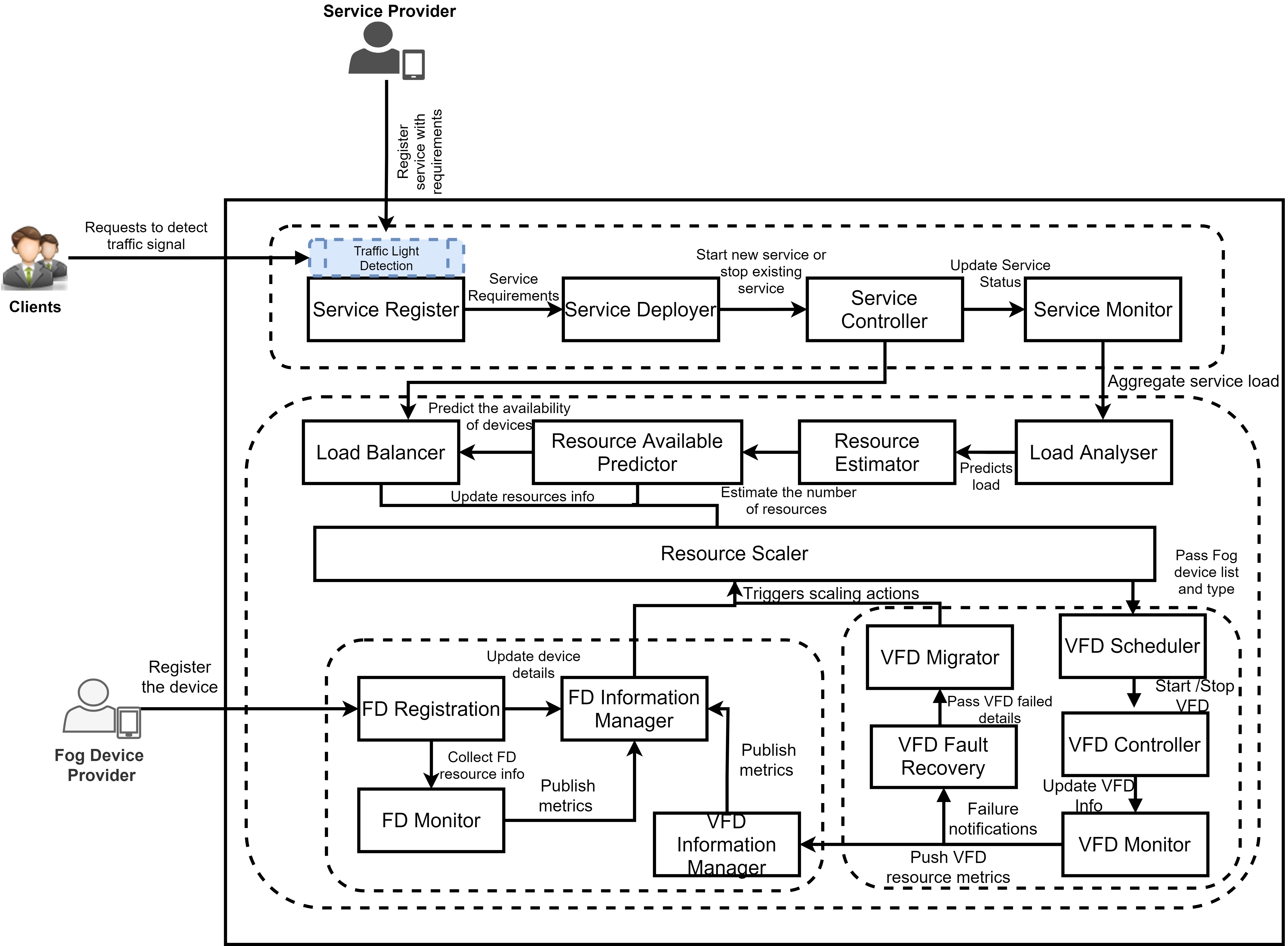}
	\caption{A comprehensive view of a FaaS Workflow.}
	\label{faasflow}
\end{figure*}
The Service Manager is responsible for registering, deploying, monitoring and controlling the services.\\
\textbf{Service Register:} The service register is responsible for accepting the service request and maintains the information about the service requirements.\\
\textbf{Service Deployer:} s responsible for passing on the decision of deploying new services to the service controller.\\
\textbf{Service Controller:}  :	is 	responsible 	for 	controlling 	the services.\\
\textbf{Service Monitor:} The service monitor updates the status of the service requested by the client periodically and updates the information at the same time.\\

The service considered in this work collects the data from the IoT devices and processes them in the virtual Fog devices and sends the results back to the server. In this section, we discuss examples of use cases that we consider validate the FaaS. An example of an application service is a traffic signal alerting service. Although this type of application or service already exists in the Cloud, we deployed it in Fog devices in FaaS to increase the latency of the applications.
To be realistic in real-world situations, the processing time for this type of application must be in the order of a few seconds, so that the users can respond rapidly. Thus, we implement these applications in Fog architecture and deploy them in Fog devices to improve their latency.
Note that our contribution is not the applications themselves; rather, we use these applications to validate the platform.

A Traffic Signal Alerting Service has three modules (i) a
tTraffic signal input capturer (ii) a tTraffic signal detector, and (iii) a uUser aAlerting system.  The traffic signal input capturer module is responsible for capturing the images at particular intervals of time and send sending those images to the Traffic traffic signal detector module. The traffic signal detector module is responsible for receiving the inputs from the traffic signal image capturer, and identify identifying the signal colour from the images by performing image analysis (Using Open CV programs) and send sending the information to the userr alerting system. The user alerting system is responsible for alerting the user, based on the information provided by the traffic detector module.


\subsection{FaaS WorkFlow}

As depicted in Figure \ref{faasflow},  the Fog device provider registers the Fog device with the FD registration and FD registration specifies the monitoring interval. Then the FD monitor service will collect the resource information from the FDs and push the resource utilisation metrics to the FD information manager. The service provider registers the service with the service register, along with the requirements. These requirements are sent to the service deployer to deploy the service in the virtual Fog devices. The service deployer requests the service controller to process the request based on the requirements specified during the registration. The service controller sends the allocated request to the resource scaler. The resource scaler passes the FD list and its type to the VFD scheduler, based on the information collected from the FD information manager. The VFD scheduler sends a request to the VFD controller, bypassing the information about the time and duration that VFD should run in the particular FDS. The VFD controller creates VFD on the FD and deploys the service. This information will be updated to the VFD monitor. The VFD monitor keeps tracks of all the VFD resource information. Based on the system load specified and on the service execution time, the FD monitor and VFD monitor will trigger an event for scaling services. Based on this event, it will notify the resource scaler to allocate or de-allocate the resources based on information collected. The VFD monitor also sends the information about the failed VFD details to the VFD controller. The VFD controller forwards the information to the VFD migrator and the VFD migrator will migrate all VFDs along with the services to other FDs through the resource scaler.

\section{Performance Evaluation}
\label{sec:7performanceeval}

This section describes the experimental environment, the evaluation parameters used to evaluate our proposed approach, the results and the discussion.

\subsection{Experimental Environment}

To emulate the Fog environment, we used two Raspberry PI devices, five desktop machines and one laptop. The configuration of these devices is given in Table \ref{Fogdev}. In order to imitate the behaviour of public Fog nodes, we are running different sample applications to increase the system load. In addition, we are disconnecting and connecting the devices automatically through a public Fog dynamic behavioural module. We use both wired and wireless connections to communicate between the devices.

\begin{figure}[h]
	\centering
	\includegraphics[width=0.5\textwidth]{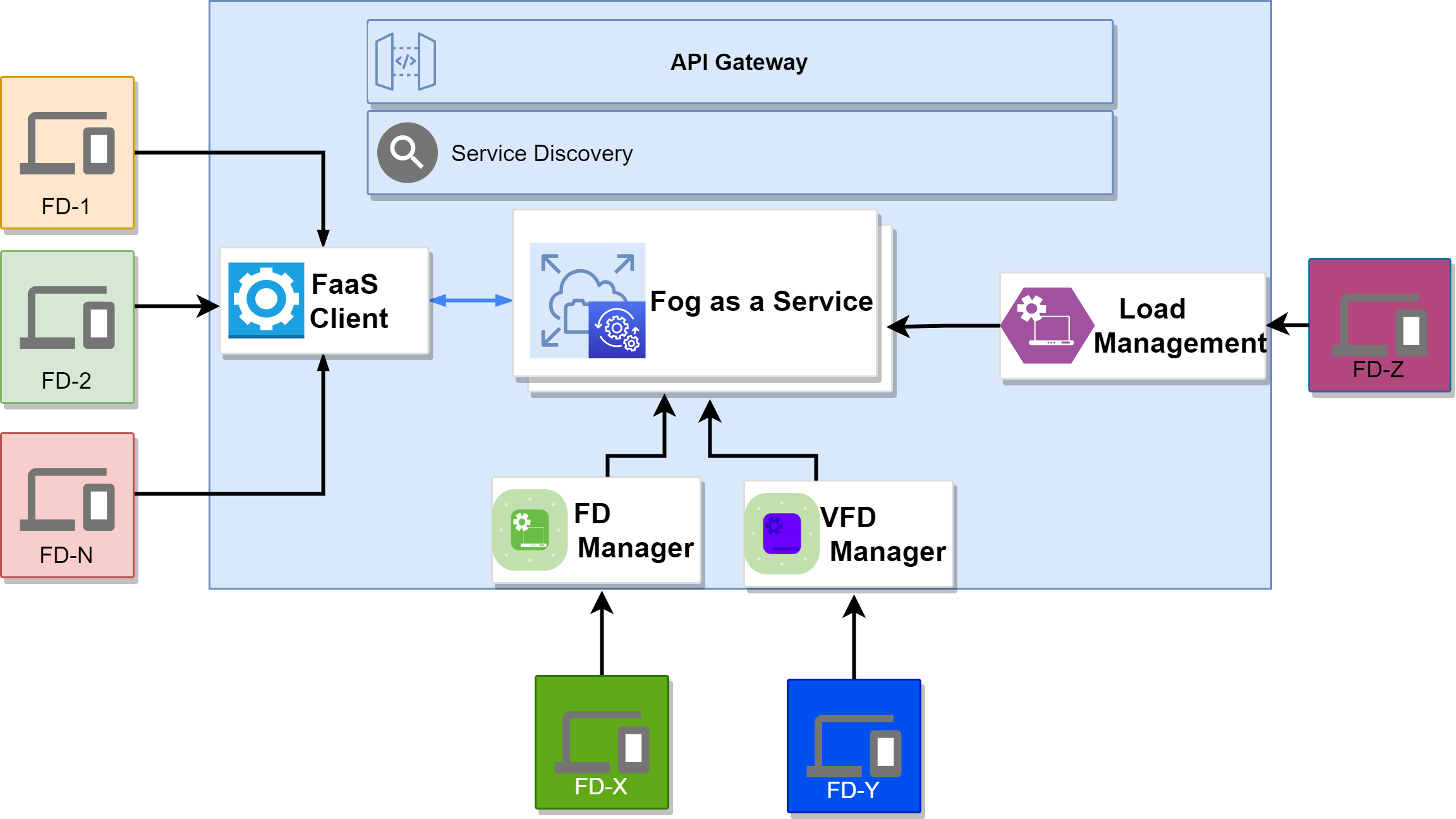}
	\caption{FaaS Deployment}
	\label{hist}
\end{figure}

FaaS has been developed with Java, Springboot, JPA, Spring
Rest, Discovery and each micro-service has been deployed in different Fog devices; these micro-services have interacted with the rest of the APIs.

The proposed algorithm is evaluated with scaling approaches called Scale-X in two scenarios. In the Scale-X approach, whenever the virtual Fog devices have reached the X\% of CPU and RAM, checks the host Fog device resource utilisation and, if the percentage of the host FD resource utilisation is less than X\%, then it will perform vertical scaling by creating one more VFD on the same FD. If the host FD resource utilisation is greater than or equal to X\%, then it will perform horizontal scaling by creating a new VFD on the other available host FD. In Scenario 1, horizontal scaling does not perform any migration of the VFD because the VFD will be created on the FD which has the image available. In Scenario 2, horizontal scaling requires the migration of VFD to start the new VFD in another FD. We adapted Scale-X percentages as maximum threshold 80 and 90 to scale, minimum threshold less than 30 to de-scale from ~\cite{6838340}--\cite{casalicchio2019study}.  We did not investigated less than Scale-80, because the most of the devices resources will be under utilisation and inefficient for the larger nodes network~\cite{10.1145/3148149}.

\begin{table}[]
	\caption{Fog device configuration details}
	\label{Fogdev}
\begin{tabular}{p{1.6cm}|p{6.5cm}}
\textbf{Fog Device Type} & \textbf{Configuration} \\
Desktop &  Intel core 2 Duo CPU E7500@2.93GHZX2 with memory 4 GB and Storage 150 GB with OS type Ubuntu 64 bit
    \\
    Raspberry PI \footnote{https://www.raspberrypi.org/products/raspberry-pi-3-model-b/} & CPU: Quad Core 1.2GHz 64bit; 
 RAM: 1GB; OS: Rasbian
    \\
Banana PI \footnote{http://www.banana-pi.org/m4.html} & CPU:
Realtek RTD1395 ARM Cortex-A53 Quad-Core 64 Bit; RAM: 2 GB DDR4 memory,
\end{tabular}
\end{table}

\subsection{Performance metrics}
We consider the following parameters to evaluate our proposed framework~\cite{ASLANPOUR2020100273}:
\begin{enumerate}
    \item Unmet requests percentage: the number of requests that fail to serve the requests on time by the total number of requests processed. We considered that the application deadline requirement to be 100 ms - 500 ms. If the response time is more than the deadline requirement, we considered it to be an unmet request.
    \item 	Average response time of unmet requests: the time taken to process unmet requests to the total number of requests.
    \item 	Average response time: the time taken to process each request by total number of requests.
\end{enumerate}

\section{Results and Discussion}
\label{sec:7results}

This section discusses the importance of our proposed technique by analysing its efficiency compared with Scale-X techniques in two different scenarios. 

\begin{figure}[h]
	\centering
	\includegraphics[width=0.5\textwidth]{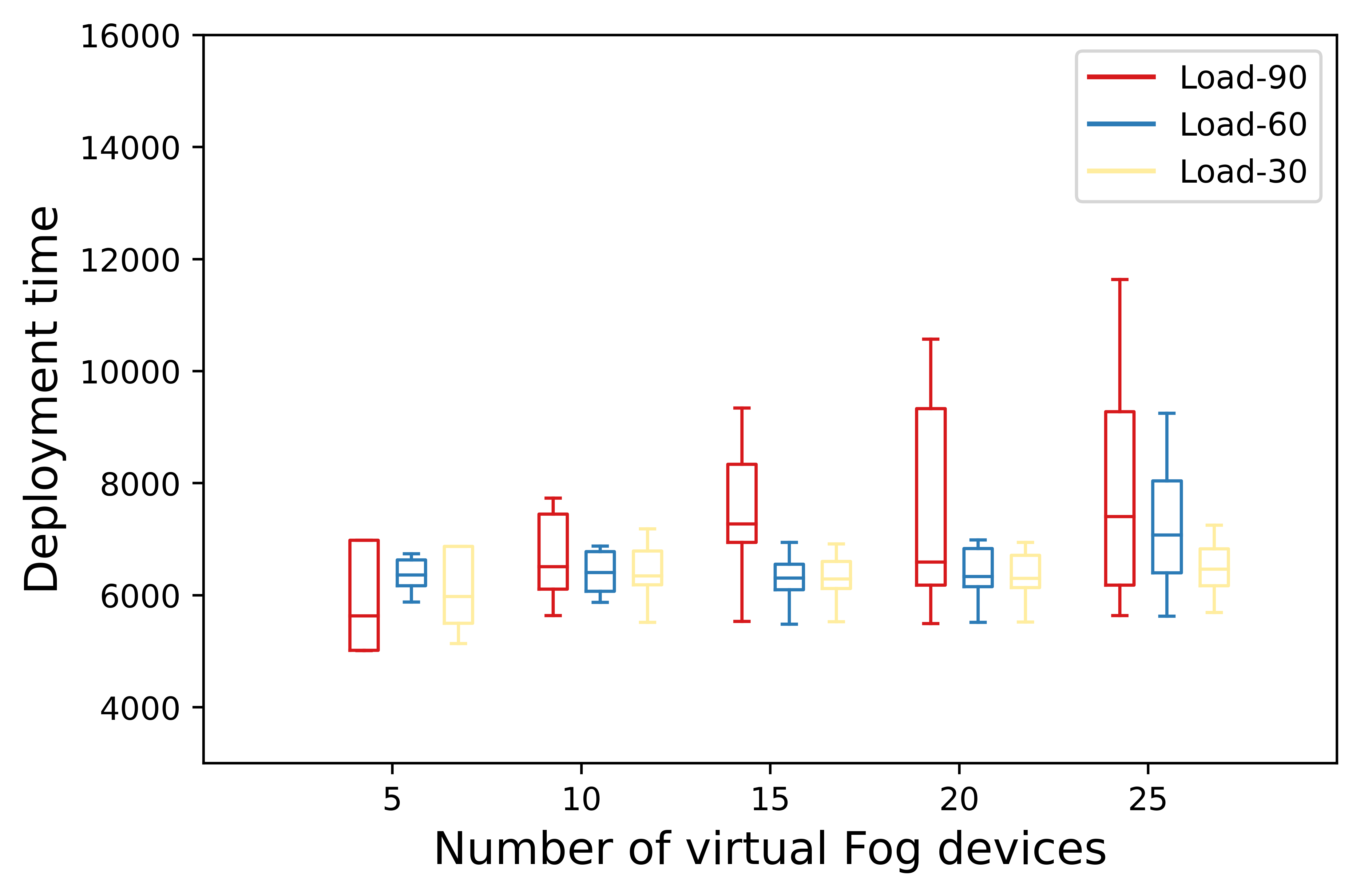}
	\caption{Deployment time of Virtual Fog devices }
	\label{deploy}
\end{figure}

\begin{figure}[h]
	\centering
	\includegraphics[width=0.5\textwidth]{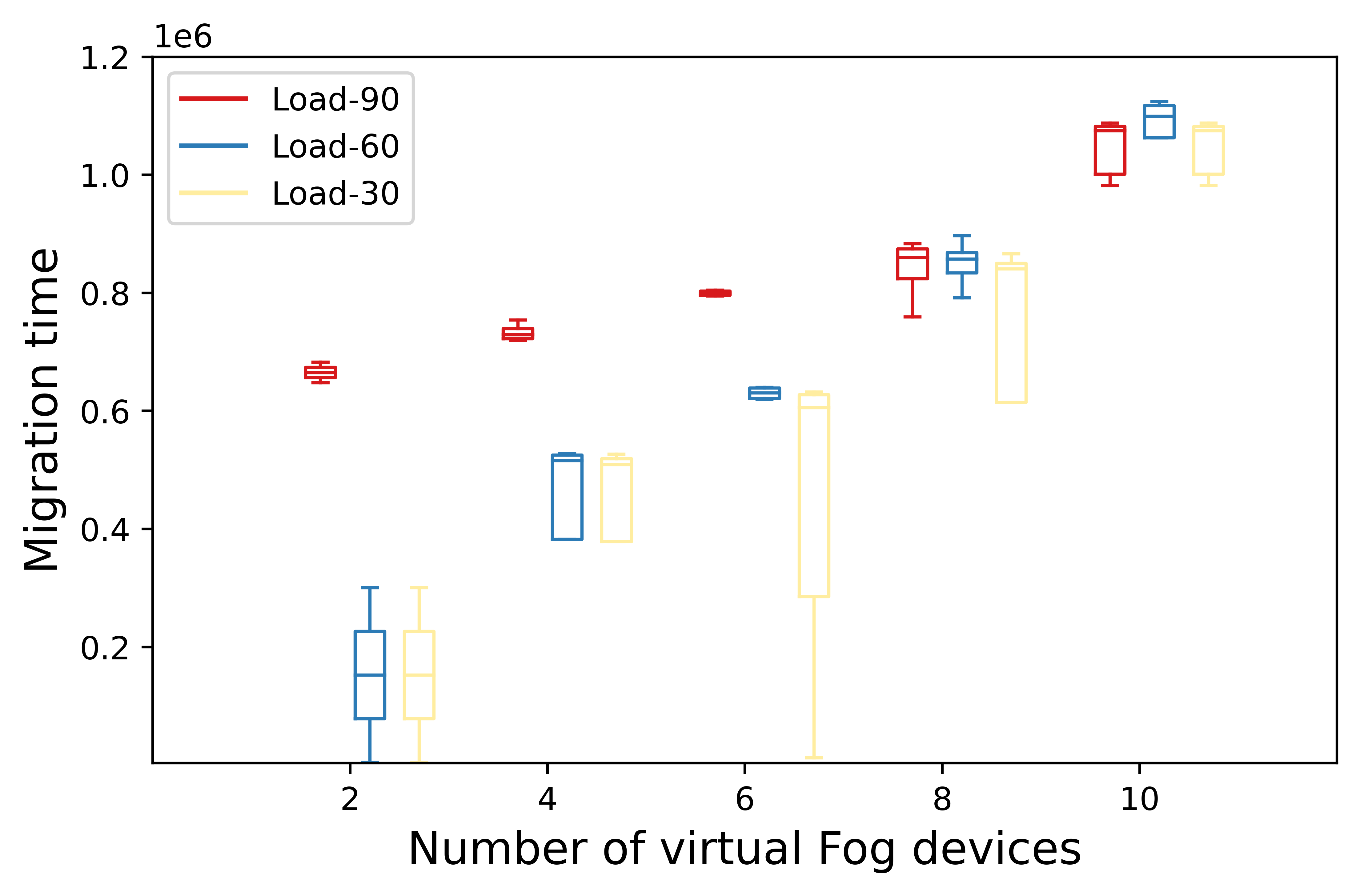}
	\caption{Migration time of Virtual Fog devices}
	\label{migration}
\end{figure}

Figures \ref{deploy} and \ref{migration} summarizes the performance of the different CPU load types. Figures \ref{deploy} shows the time taken to deploy the number of virtual Fog devices under different CPU loads. In Load-90, the CPU load varies from 60\% to 90\% of CPU utilisation; the deployment time increases from 5 VFD to 25 VFD due to the high CPU load. In Load60, the deployment time is higher than in Load-30. This is expected due to the high CPU Load.

Figures \ref{deploy} indicates the time taken to migrate the number of virtual Fog devices to different Fog devices under different CPU loads. When the CPU load varies from 60\% to 90\% of CPU utilisation, the migration time increases from 2 VFD to 10 VFD, compared with Load-60 and Load-30. This is expected due to the high CPU load. The Load-30 deployment time of different virtual Fog devices is less than Load-60. This is expected due to the low CPU load.
\newline

\subsection{No migration required when performing horizontal scaling}

\begin{figure}[h]
	\centering
	\includegraphics[width=0.5\textwidth]{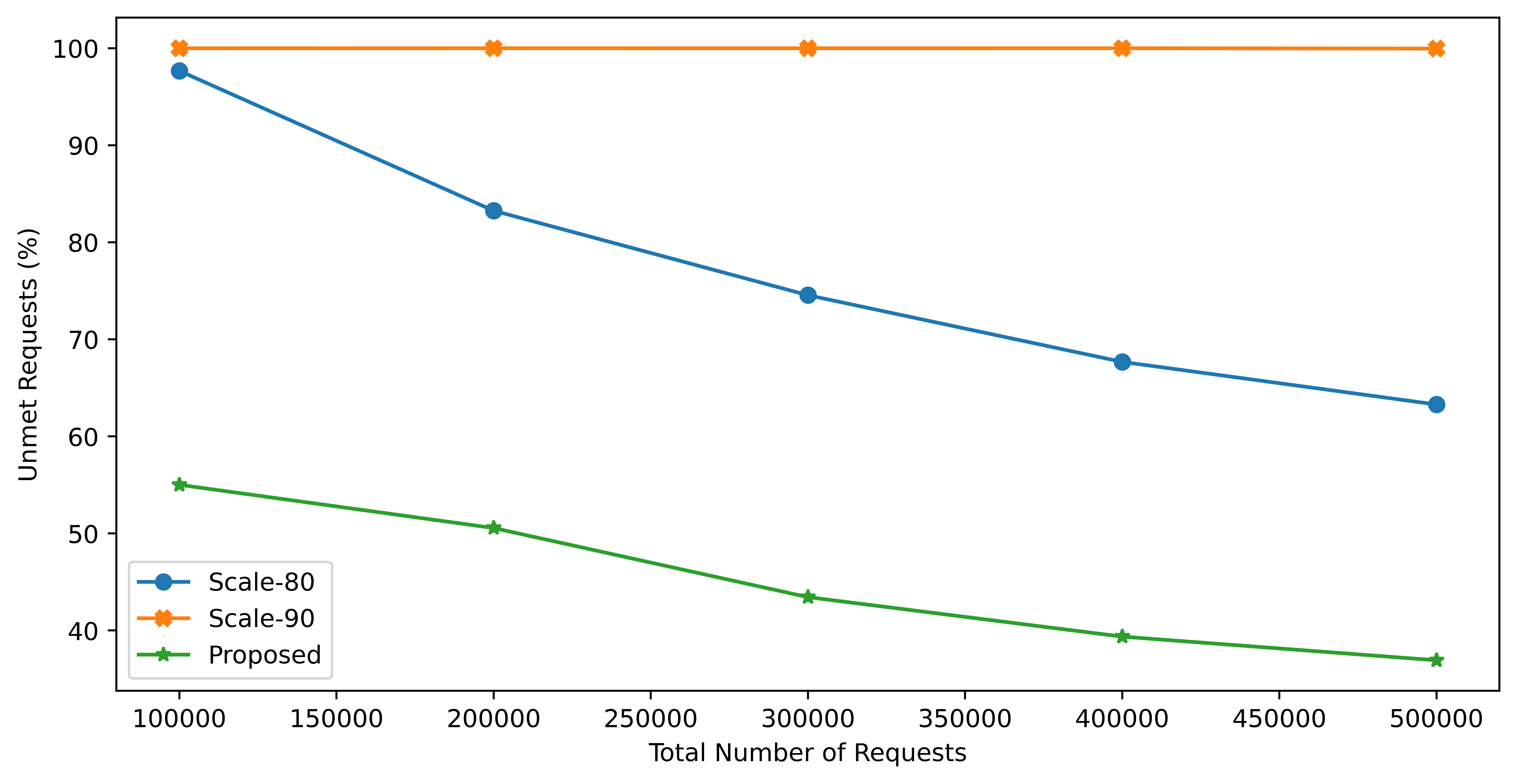}
	\caption{Unmet requests percentages of different application requests (No migration of VFD)}
	\label{unmetno}
\end{figure}

\begin{figure}[h]
	\centering
	\includegraphics[width=0.5\textwidth]{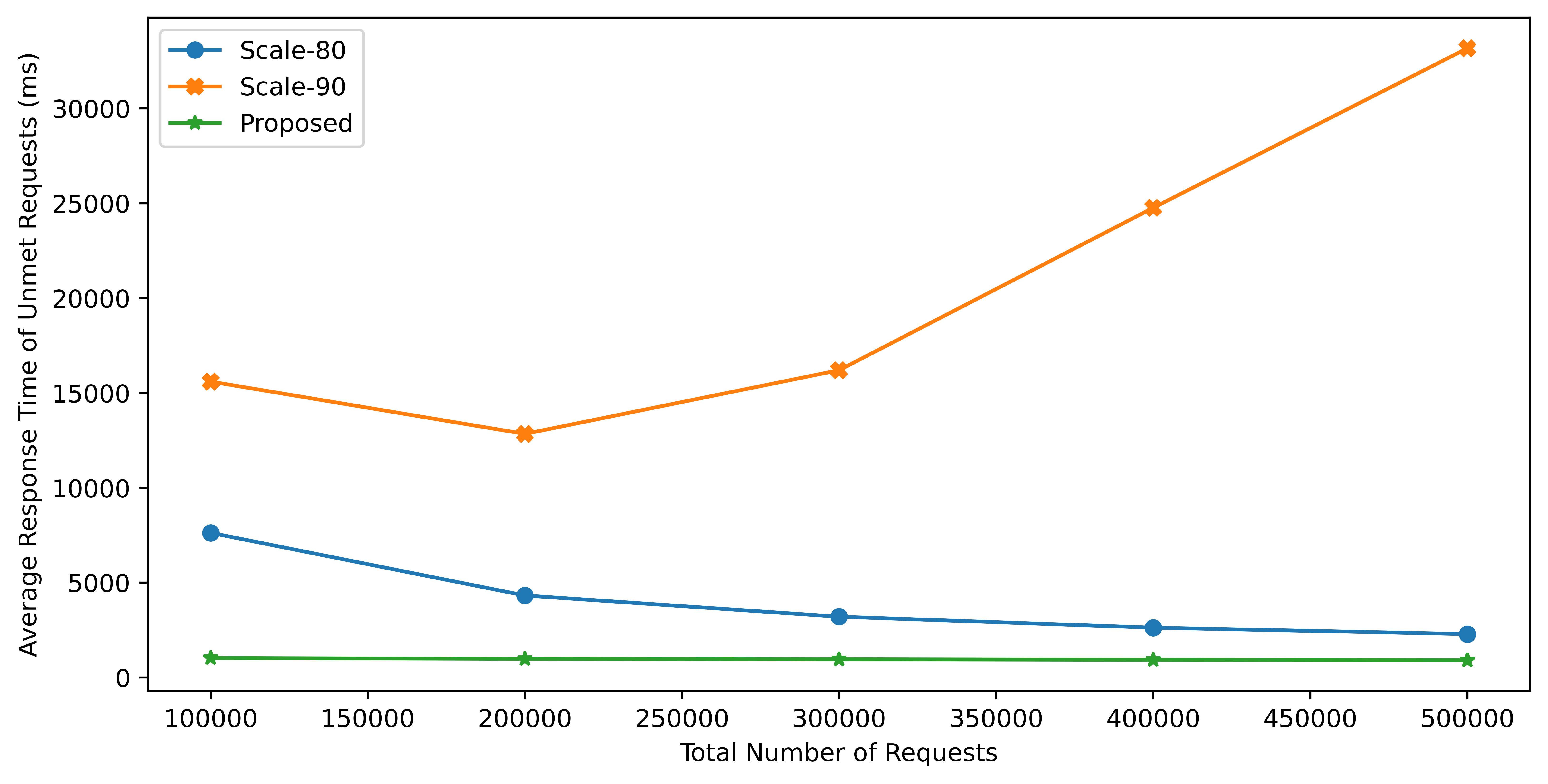}
	\caption{Unmet requests average response time of different application requests (No migration of VFD)}
	\label{unmetresno}
\end{figure}

\begin{figure}[h]
	\centering
	\includegraphics[width=0.5\textwidth]{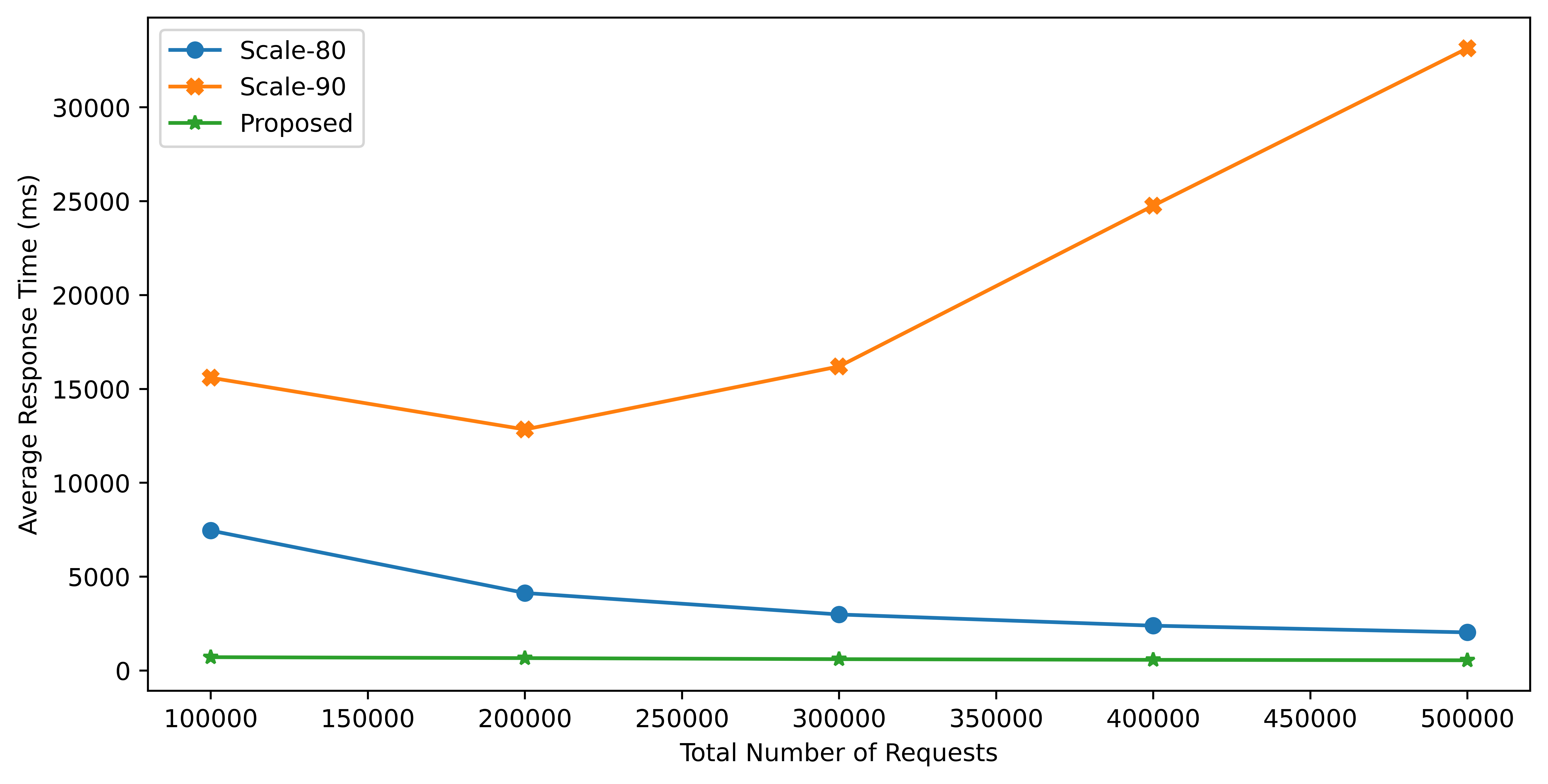}
	\caption{Average response time of different application requests (No migration of VFD)}
	\label{avgresno}
\end{figure}

Figures \ref{unmetno}--\ref{avgresno}, represent the percentage of unmet requests, the average response time of unmet requests and the average response time of Scenario 1, as described above. 

Figure \ref{unmetno} shows the efficiency of the proposed algorithm in terms of the percentage of the unmet requests in comparison with the Scale-80 and Scale-90 approaches when increasing the number of requests. The proposed mechanism has fewer unmet requests compared with the Scale-80 and Scale-90 approach because the proposed algorithm predicts the number of application requests, and creates and allocates the virtual Fog devices to deploy the services, by predicting the host utilisation for the duration of the demand. However, the unmet requests percentage in Scale-80 and Scale-90 is more due to the greater number of requests accommodated in each VFD with which the response time was increased and does not meet the time requirement of the application. Moreover, there is a decreasing trend in both the proposed and the Scale-80 techniques because, in the increasing number of FaaS instances, the requests are balanced between the VFDs.

The proposed algorithm has the lowest average response times and the average response time of unmet requests is shown in Figures \ref{unmetresno} and \ref{avgresno}. The average response time of total requests in the proposed approach is varied between 543ms to 710 ms; the Scale-80 average response time is varied between 2030 ms to 7454 ms. The Scale-90 average response time is varied between 15597 ms to 33148 ms. This is because the number of VFDs required is predicted before in our approach and deploys the FaaS on the predicted FDs, based on the information of the available resources, and distributes the request load between them.

However, the proposed approach has low resource utilisation compared with other approaches. This is because the idle resources in the peak periods are not enough to process the requests in a timely manner.

\subsection{Migration required when performing horizontal scaling}

\begin{figure}[h]
	\centering
	\includegraphics[width=0.5\textwidth]{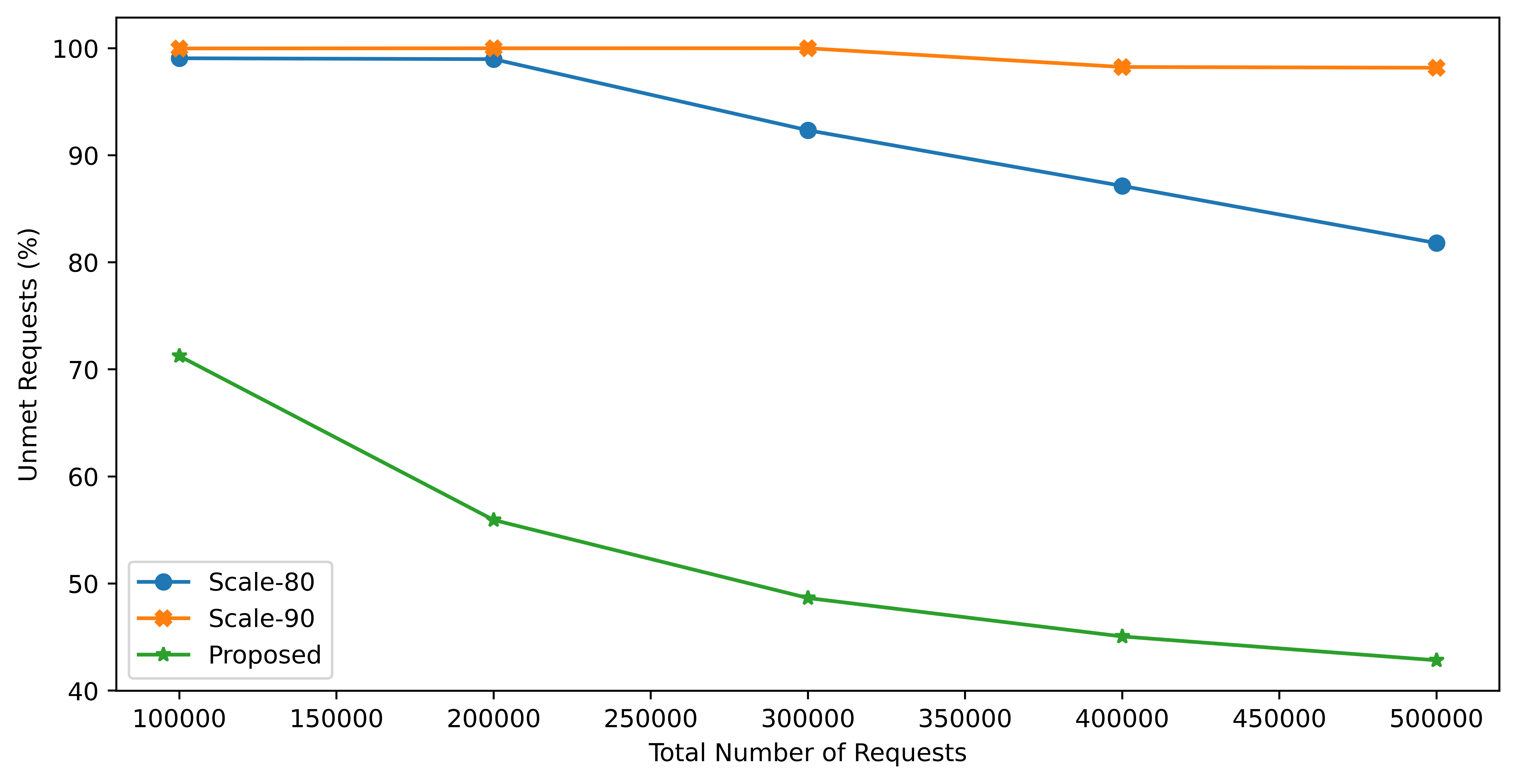}
	\caption{Unmet Requests percentages of different application requests (Migration of VFD)}
	\label{unmet}
\end{figure}

\begin{figure}[h]
	\centering
	\includegraphics[width=0.5\textwidth]{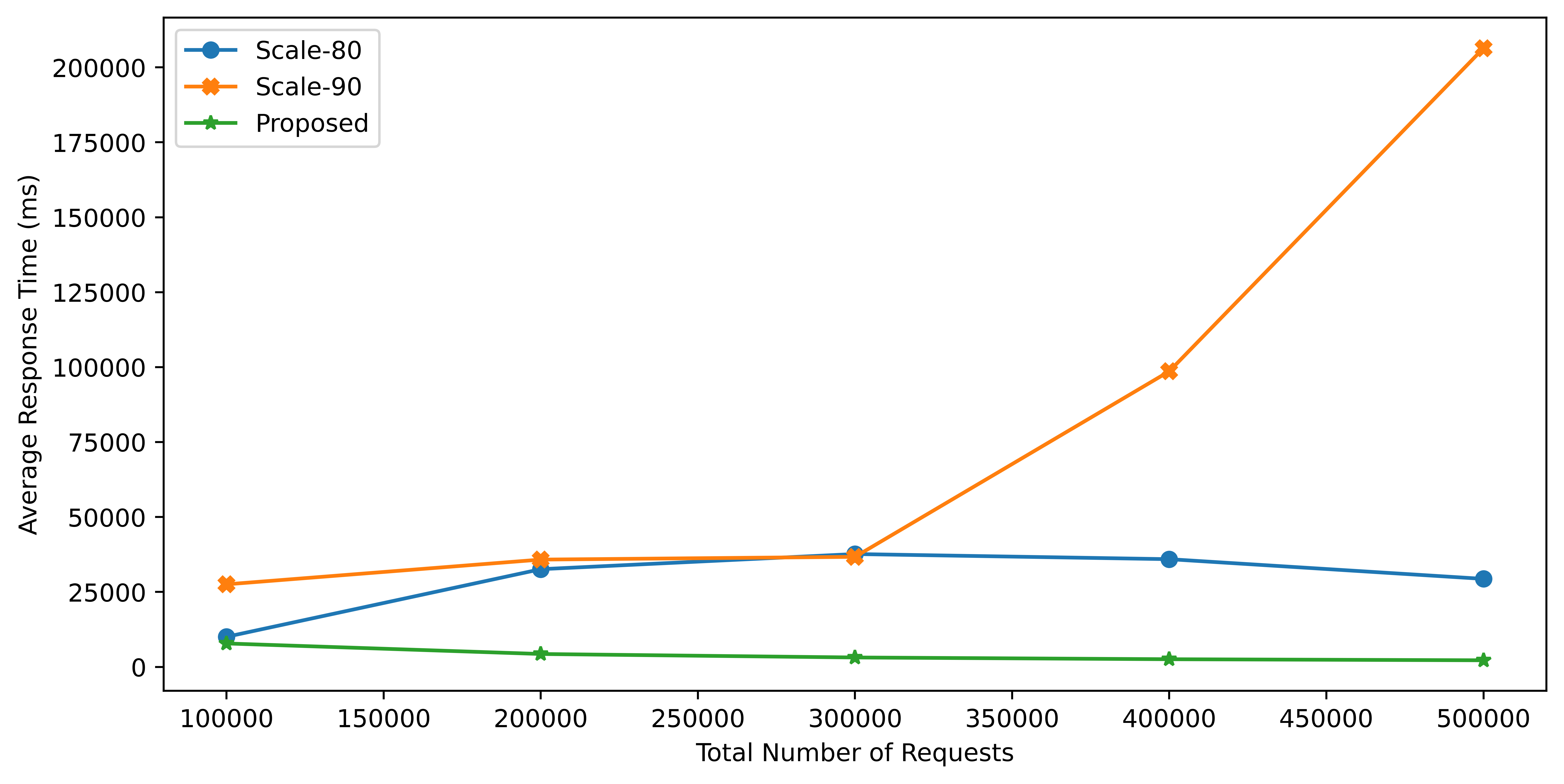}
	\caption{Unmet Requests average response time of different application requests (Migration of VFD)}
	\label{unmetres}
\end{figure}

\begin{figure}[h]
	\centering
	\includegraphics[width=0.5\textwidth]{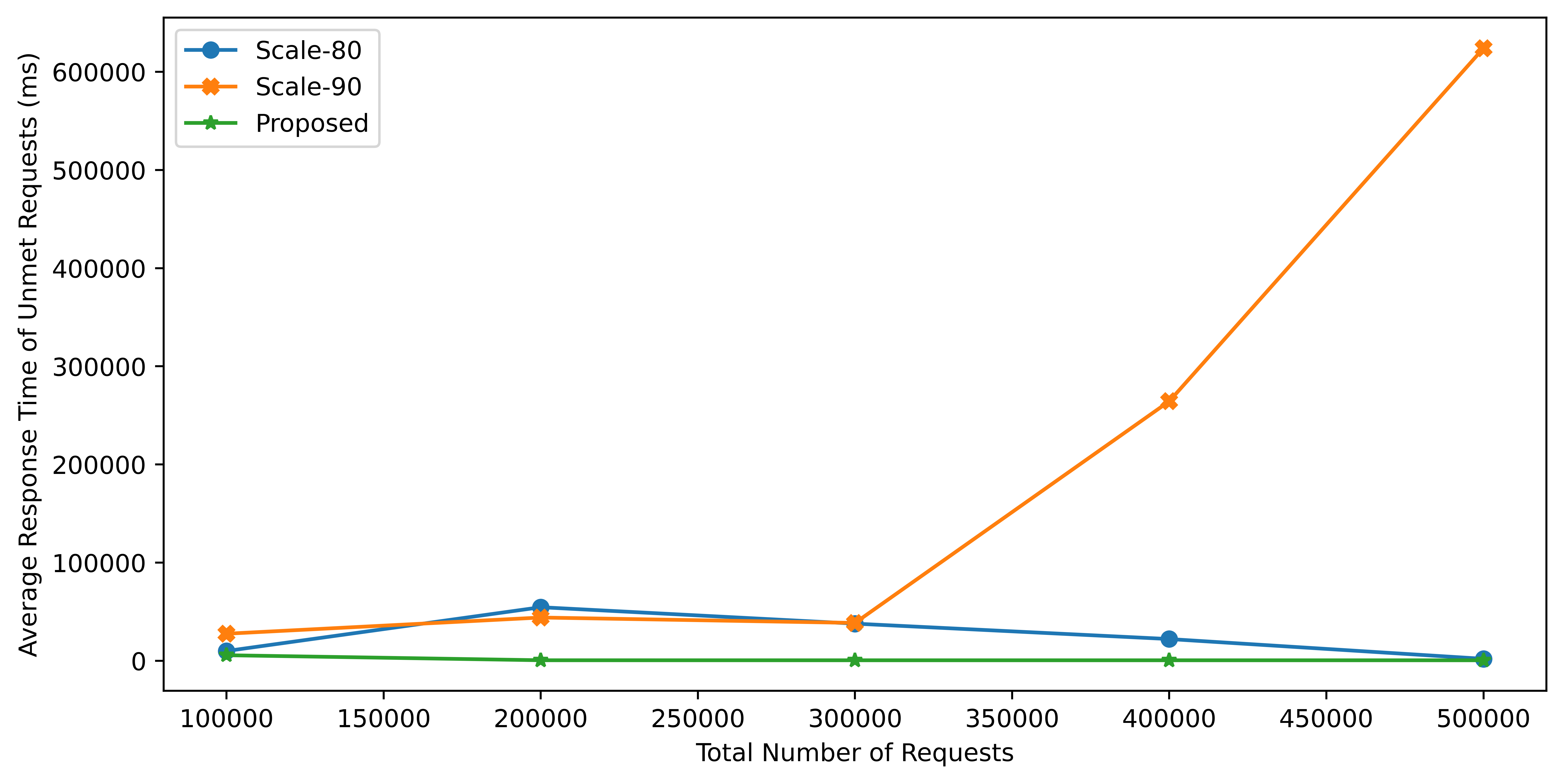}
	\caption{Average response time of different application requests (Migration of VFD)}
	\label{avgres}
\end{figure}

Figures \ref{unmet}--\ref{avgres} show the experiment results of Scenario 2. Our proposed approach has the lowest unmet request percentage compared with the other approaches, as reported in Figure~\ref{unmet}. 
Moreover, the number of unmet requests is very high in Scale-80 and Scale-90 during the first 100,000 requests because the resources of FD are utilised more in creating the snapshots and migrating, due to which the response time of the requests has been increased.

In Figure \ref{avgres}, there is a decreasing trend in both the proposed and Scale-80 techniques because in the increasing number of FaaS instances, the requests are balanced between VFDs. However, Scale-90 has almost the same trend this is because of overloaded requests in each VFD.
Due to the increase in the number of requests and un-reached the targeted resource utilisation leads to accommodate the more number of requests to the FaaS. Thus, average response time is an increase in Scale-90 as shown in the Figures \ref{unmetres} and \ref{avgres}. 

The proposed approach has the lowest average time compared with both other approaches. This is because, in our proposed approach, the number of FaaS deployed in the number of VFDs is determined based on the requirement of the application (deadline of the request). Also, the selection of FDs is based on the prediction of future availability of the resources, based on the current resource utilisation. However, the other approaches will only create a new FaaS instance when they reach the targeted resource utilisation.
\section{Conclusions and Future Work}
\label{sec:7conclusion}

In this paper, we evaluated a system based on the proposed FaaS framework for the dynamic deployment of FaaS services on virtual Fog devices in the Fog computing environment. We proposed a dynamic resource allocation approach by exploiting device availability information. Our proposed approach is implemented and evaluated in a real environment. The SARIMA model used in our approach helps to predict future application requests. The results show that our proposed approach tends to leave fewer requests unmet compared with the other approaches. It maximises the satisfied service requests by an average of 1.9 times in Scenario 1 and an average of 1.8 times in Scenario 2.
Finally, as part of future work, we aim to improve the device registration process and plan to create dynamic clusters to support Big data applications. We also have a plan to integrate Smart Software Defined Network (SSDN) to automatically control the size of the data between the clusters, in order to improve the performance of the Big data streaming applications.

\bibliographystyle{IEEEtran}
 \bibliography{Reference.bib}

\end{document}